# Single chip 1 Tb/s optical transmitter with inverse designed input and output couplers

Kaisarbek Omirzakhov[1], Ali Pirmoradi[1], Geun Ho Ahn[2], Amirreza Shoobi[1], Han Hao[1], Jelena Vučković[2], and Firooz Aflatouni[1*]

[1]Department of Electrical and Systems Engineering, University of Pennsylvania, Philadelphia, PA, USA
[2]E. L. Ginzton Laboratory, Stanford University, Stanford, CA 94305, USA
*firooz@seas.upenn.edu

**Optical interconnects are essential for data centers and AI systems. Given the limited energy production, ultra-low energy and dense optical interconnects are required to support the exponential growth of AI systems. Here we report the demonstration of a monolithically integrated optical transmitter where use of power efficient architecture and devices such as capacitive tuning of optical structures at zero static power consumption and efficient and wideband inverse designed grating couplers enable implementation of a 32-channel transmitter chip based on wavelength-division multiplexing achieving a record modulation energy efficiency of 32 fJ/bit at 5 Gb/s/channel and 106 fJ/bit at 32 Gb/s/channel, which includes the tuning of optical devices. Furthermore, a record bandwidth density of 8 Tb/s/mm$^2$ for a bit-error-rate of 10$^{-12}$, while all channels are simultaneously operating with an aggregate data-rate of 1.024 Tb/s, was achieved. The system utilizes 16 carrier wavelengths in the optical C band. The pseudo-random-bit-streams are electrically generated on-chip and used to drive individually wavelength-stabilized 2-section p-n-capacitive micro-ring modulators using integrated energy-efficient high-swing electrical drivers. The low-loss inverse designed grating couplers have -1-dB bandwidth of 25 nm. The chip, implemented using GlobalFoundries 45CLO CMOS SOI process within a footprint of 1.55 mm × 1.11 mm, concurrently achieves the highest aggregate data-rate, the highest energy efficiency and the highest bandwidth density for a multi-channel high date-rate optical transmitter chip reported to date.**

High performance computing systems and data centers are facing unprecedented demand due to the growth of data-intensive applications such as generative artificial intelligence (AI). Given the limited global energy production, the exponentially growing energy demand of such systems can slow down their scalability. Meanwhile, existing infrastructure, which is often optimized for traditional computing tasks, undergo continuous optimization to support the ever-increasing compute demand. Modern AI models, especially large deep neural networks, require massive parallel computing units such as graphics processing units (GPUs), tensor processing units (TPUs) and high bandwidth memory modules[1,2,3] and face significant bottlenecks due to the limited throughput and energy efficiency of the high-speed local interconnects between modules, both at the scale-up and scale-out levels[4]. As such systems are expected to continue growing[5], addressing the bottleneck in energy-efficient data movement becomes a critical challenge.

Optical links have emerged as a viable solution for high-bandwidth and low-loss, long-reach interconnects and play a crucial role in advancement of data centers and long-haul communication networks[6]. Industry-standard optical pluggable modules can be seamlessly integrated with existing infrastructures while enhancing overall data throughput[7] and have evolved to support data-rates of 800 Gb/s and beyond[8]. However, despite recent advancements, the energy efficiency, bandwidth density and latency of current pluggable optical interconnects need to be further improved and new innovations in optical device technology, system architecture and packaging are needed to meet the growing needs of modern computing infrastructure[9,10,11]. Recent advances in silicon photonics have paved the way for compact, high-bandwidth, low-latency and energy-efficient interconnects, primarily realized through either co-packaged (hybrid) or monolithic integration[12-22]. Both integration approaches have been extensively explored for optical links, each offering distinct advantages and trade-offs. Hybrid integration leverages separate electronic and photonic chips, fabricated using state-of-the-art processes and vertically stacked often through micro-bumps and flip-chip bonding. While this method enables the use of high-performance electronic and photonic components, the associated packaging parasitics could limit energy efficiency and operational bandwidth.

Monolithic integration minimizes parasitic components and reduces packaging complexity, resulting in improved energy efficiency and higher data-rates. However, the performance of on-chip electronic and photonic devices may fall short of the performance achieved by state-of-the-art components fabricated through specialized CMOS-only or photonics-only processes.

Here we report the demonstration of a monolithic 32-channel optical transmitter chip that utilizes wavelength division multiplexing (WDM) and the non-return to zero (NRZ) modulation scheme to achieve an aggregate data-rate of 1.024 Tb/s using a single-mode input optical fiber, delivering 16 optical carriers with carrier-to-carrier spacing of 200 GHz to the chip, and four output (link) fibers. Optical carriers are split on chip and modulated at 32 Gb/s with different bit-streams. Reduced parasitics between electronic circuits and photonic devices enabled by monolithic integration as well as use of energy-efficient and low-loss techniques and designs such as capacitive tuning of optical structures at zero static power consumption, low-power electronic circuits, and wideband and efficient inverse designed grating couplers results in a record system energy efficiency. The modulators are novel 2-section p-n-capacitive micro-ring modulators (MRM) with energy efficiency of 106 fJ/bit and 32 fJ/bit at 32 Gb/s and 5 Gb/s, respectively. The chip achieves a bit-error-rate of less than $10^{-12}$ while operating at 1.024 Tb/s (32 channels concurrently operating at 32 Gb/s/channel). Furthermore, compact design for electronic and photonic devices results in a record areal bandwidth density of 8.0 Tb/s/mm$^2$. The input and output grating couplers are implemented using inverse design approach to support a -1-dB optical bandwidth of 25 nm, needed to support 16 carriers with 1.6 nm spacing, while achieving a higher coupling efficiency compared to the standard grating couplers.

The chip, implemented using the GlobalFoundries GF45CLO process (a 45 nm CMOS SOI process)[23], paves the way for realization of low-power high data-rate optical links for future data centers.

## Results

Figure 1 shows the block diagram of the implemented monolithically integrated 32-channel transmitter chip. A single-mode fiber delivers 16 optical carriers with 1.6 nm carrier-to-carrier spacing in the optical C band to the chip, where they are coupled into the chip through the input inverse designed grating coupler and routed to the input Mach-Zehnder interferometer (MZI). The MZI de-multiplexes the even and odd carriers at its two output waveguides. Each output of the MZI demultiplexer is equally split using a Y-junction splitter and each of the 4 outputs of the two Y-junctions serves as the common bus waveguide for 8-cascaded 2-section p-n-capacitive MRMs. On-chip test data generator units generate different 32 Gb/s pseudo-random bit sequences that are used to modulate the carriers within the 32 channels by driving the p-n sections of the MRMs using low power modulator drivers. The modulated outputs of 8 cascaded MRMs on the same bus waveguide are coupled out of the chip using an inverse designed grating coupler. To maintain wavelength alignment, the wavelength locking and carrier tracking system wavelength-locks each MRM to its corresponding target optical carrier. MRMs in channels 1-16 are designed to modulate carriers with an odd wavelength number (i.e. $\lambda_1, \lambda_3, \ldots \lambda_{15}$) and MRMs in channels 17-32 are designed to modulate carriers with an even wavelength number (i.e. $\lambda_2, \lambda_4, \ldots \lambda_{16}$).

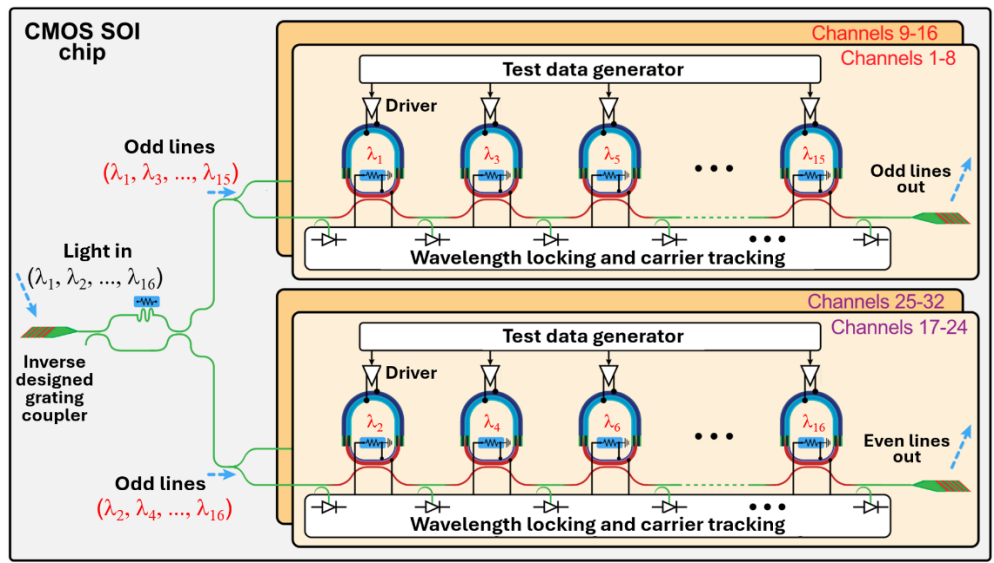

**Fig. 1 | System architecture.** Block diagram of the implemented monolithically integrated 32-channel transmitter for 16 wavelengths based on 2-section p-n-capacitive micro-ring modulators and inverse designed input/output grating couplers.

Given the required input optical bandwidth, which spans 16 optical carriers with 1.6 nm carrier-to-carrier spacing, ideally, the free spectral range (FSR) of each MRM should be set to be larger than the input optical bandwidth and by cascading such MRMs with varying FSRs on a bus waveguide, target carriers can be selected and modulated with the corresponding bit-streams[24]. Despite offering a straightforward frequency planning, this approach poses certain practical challenges for the MRM design as the large required FSR (and hence a rather small MRM diameter) results in increased optical loss and reduced modulation efficiency due to reduced MRM quality factor.

Another approach for selecting and modulating 16 optical carriers is to use MRMs with FSRs smaller than the input optical bandwidth. In this case, while using MRMs with larger perimeters results in a lower optical loss and a higher MRM quality factor and modulation efficiency[25, 26], a careful FSR design is required to limit the crosstalk between the channels. In this work, besides optimizing the FSR of the MRMs, the crosstalk is further reduced by separating the even and odd wavelength using the input

1:2 MZI-based demultiplexer, effectively doubling the carrier-to-carrier spacing on each waveguide bus (which contains eight cascaded MRMs). Additionally, in this work the number of modulator channels are set to 32, which requires sharing 1 wavelength between two MRMs. This is achieved by using Y-junctions to equally split the power of a set of eight carriers between two sets of bus waveguides (each feeding eight MRMs). The details of the design of the system wavelength response and carrier demultiplexing are included in Method section and Extended Data Fig. 1.

**Inverse designed grating couplers**

Given the carrier-to-carrier spacing, the required input optical bandwidth linearly increases with the number of optical carriers. This input optical bandwidth sets the minimum bandwidth requirement for input/output (I/O) optical interfaces. While edge couplers offer a relatively large optical bandwidth, efficient edge couplers often occupy a rather large chip area, significantly reducing the areal bandwidth density. On the other hand, despite small footprint, conventional optical grating couplers suffer from a limited bandwidth and often a larger coupling loss compared to edge couplers. In this work, the photonic inverse design method (based on a variation of the software package SPINS described in our prior works[27,28]) was used to design grating couplers with increased optical bandwidth and reduce coupling loss. To inverse design the grating coupler in the GlobalFoundries GF45CLO process three steps were considered. First, only a single layer (polysilicon) on top of the silicon slab is chosen for optimization to break the longitudinal symmetry of the grating coupler structure. Second, the objective function is framed to target the minimum required optical bandwidth while minimizing the coupling loss. Third, the optimization is constrained by the fabrication limitations (e.g. minimum feature and gap sizes) of the GlobalFoundries GF45CLO process. Figure 2a shows the structure of the implemented inverse designed grating coupler. The grating structure has a footprint of 12 µm x 12 µm. The grating structure is tapered to a 500 nm wide single-mode waveguide using a 100 µm long linear taper. The gratings in polysilicon layer are asymmetric with varying widths and gaps, which are defined by the optimization tool. Grating coupler structures were fabricated without any violations of the process design rules. Figure 2b shows the measurement results for two back-to-back inverse designed grating couplers structures, where a minimum insertion loss of 9 dB (4.5 dB per grating coupler), and the -1 dB and -3 dB bandwidths of 25 nm and 45 nm were measured, respectively. For a 1.6 nm carrier-to-carrier spacing, the implemented inverse designed grating couplers support 16 carriers with about 1 dB power variations between the carriers.

**Optical de-multiplexer and splitter**

The structure of the MZI de-multiplexer at the chip input is shown in Fig. 2c. A wideband 50/50 directional coupler[29] is used at the MZI input to equally split and route the signal into two branches. The light in the top branch is delayed using a meandered optical delay line and another wideband 50/50 directional coupler is used to combine the two branches at the MZI output. The curved directional couplers used in the MZI de-multiplexer are designed with a linear length, coupling region curve radius and gap of 50 µm, 46 µm and 200 nm, respectively. The length difference between the arms of the MZI is set to about 190 µm corresponding to an FSR of about 3.2 nm.

Figure 2d shows the measured transmission response of the implemented MZI, where an extinction ration of more than 25 dB is maintained across at least 40 nm of bandwidth. A resistive heater element, isolated using oxide trenches, was placed near the delay line, which can be used for thermal tuning in case needed. The measured $P_\pi$ of this thermal phase shifter is about 8 mW.

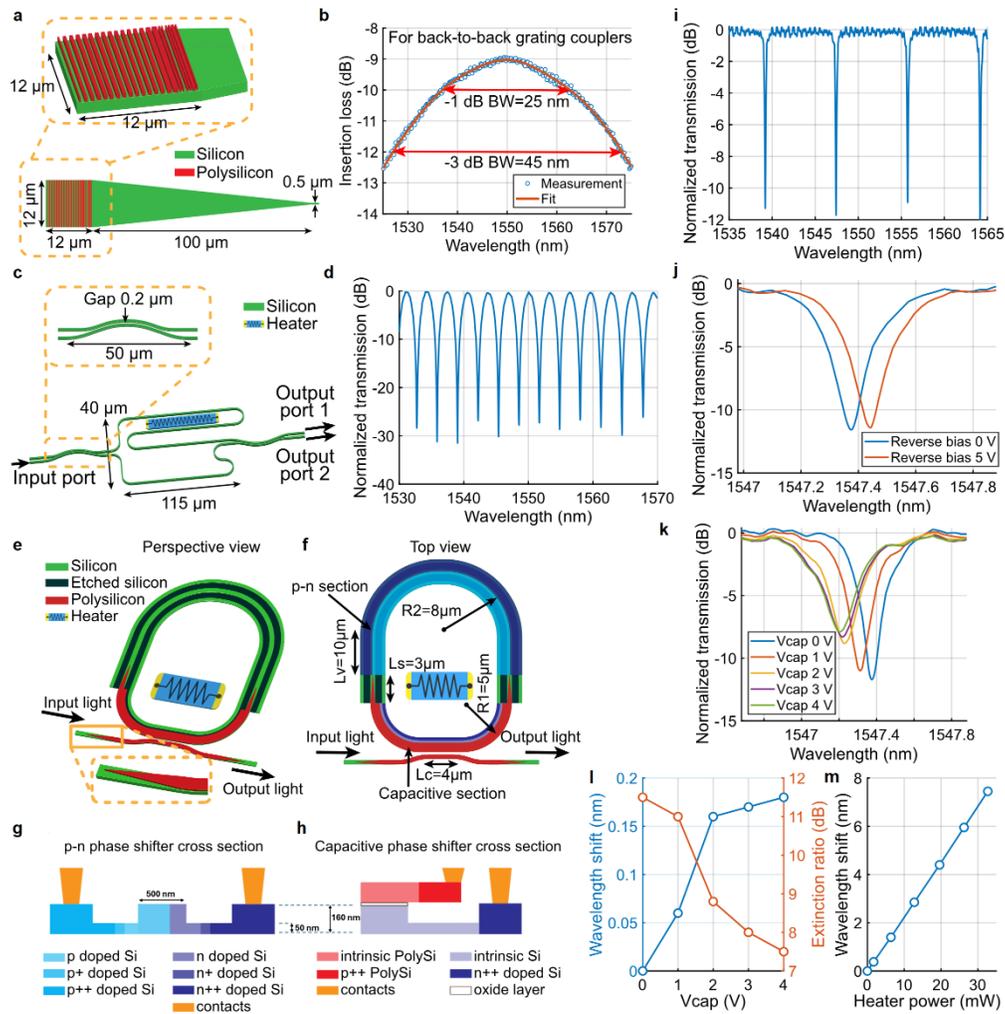

**Fig. 2 | Inverse designed grating coupler, de-multiplexer and modulator**. **a,** The implemented grating coupler structure using inverse design approach. **b,** Measured response of the back-to-back grating coupler structure. **c,** The implemented MZI structure with wideband couplers and integrated heaters. **d,** Measured normalized transmission of the MZI showing an FSR of around 3.2 nm. **e,** Perspective view of the undoped MRM and **f,** top view of the implemented MRM structures with p-n and capacitive sections. Cross sections of **g,** p-n and **h,** capacitive phase shifters. **i,** Normalized optical response of an MRM showing an FSR of about 8.4 nm. **j,** The measured tuning response of the reverse biased MRM. **k,** Measured MRM optical transmission response for different applied bias voltages across the capacitive section. **l,** Measured capacitive wavelength tuning and extinction ratio versus applied voltages across the capacitive section. **m,** Measured thermal wavelength tuning versus the heater power. Lc: length of the coupling region. Ls: length of the polysilicon taper. Lv: length of the straight region of the MRM.

**Capacitively tuned micro-ring modulators**

Micro-ring modulators offer compact footprint, high quality factor, and high bandwidth and are promising devices to enable large scale, energy efficient, and high-speed optical transmitter systems[30,31,32]. To maximize the modulation efficiency in a micro-ring modulator, different design parameters, such as coupling coefficient, ring circumference and loss need to be optimized. More details on the effect of design parameters on micro-ring resonator performance are included in the Methods section and Extended Data Fig. 2. The multi-section MRM structure in this work is inspired by the design in our prior work[12] but modified based on the frequency planning of this system, which is designed to support 32 channels given 16 input optical carriers, while minimizing the channel-to-channel crosstalk. Each MRM can be capacitively tuned and locked to the target wavelength at zero static power consumption.

The perspective view of the structure of the MRM without doping layers is shown in Fig. 2e. The top view of the implemented MRM with doping layers and the cross sections of the p-n and capacitive sections are shown in Figs. 2f, g and h, respectively. The implementation of the MRM structure does not require any foundry process modifications nor any post processing.

The p-n section is formed by laterally doping 61% of the silicon waveguide of the ring with a 66.6%-33.3% ratio. The capacitive section constitutes 31% of the ring circumference, including the ring coupling region, and is formed as a vertical stack of polysilicon-oxide-silicon (p-o-s) structure, where the oxide layer is the gate thin oxide layer used for MOSFET transistors in the GF45CLO process. Since the thickness of the thin oxide layer is much smaller than the waveguide dimensions, it does not affect the optical mode, which is confined within the vertical polysilicon-oxide-silicon structure. By applying a voltage across the capacitor, the charge profile within the capacitive modulator changes, effectively modifying the index of refraction of the p-o-s waveguide, which shifts the phase of the propagating optical wave based on the overlap between the optical mode and the altered charge profile. The FSR of each MRM is optimized to ensure the lowest possible loss and crosstalk while effectively de-multiplexing and modulating the 16 input optical carriers. Under normal operation, capacitive tuning is sufficient to compensate for the effect of process variation and lock the MRMs to the corresponding target wavelengths[29]. Besides capacitive tuning, in cases a larger tuning range is needed, a resistive heater (i.e. a lightly doped silicon structure) placed inside each MRM can be used for thermal tuning.

Figure 2i shows the normalized optical response of the designed MRM, where an extinction ratio of more than 10 dB and an FSR of about 8.4 nm is measured. The optical resonance shift for different applied reverse biased voltages across the p-n junction is shown in Fig. 2j, where a tuning efficiency of about 13 pm/V is measured.

Figure 2k shows the MRM tuning response as a function of the voltage applied across the capacitive section. The charge accumulation within the waveguide region is altered as the voltage across the capacitive section is changed such that, as shown in Fig. 2l, increasing the voltage causes the MRM resonance to shift to the smaller wavelengths (blue shift), and also increases the loss within the capacitive modulator, which reduces extinction ratio and the quality factor of the MRM. For the MRM in this work, a typical maximum resonance shift using capacitive tuning is about 180 pm at 4 V. The breakdown voltage of the capacitive section (for which the thin oxide layer between the polysilicon and silicon layers is damaged) is about 6 V.

For the unlikely case that capacitive tuning alone is insufficient for wavelength tuning of the MRMs, resistive heaters can be used. Figure 2m shows the measured thermal tuning response of an implemented MRM, where a thermal tuning efficiency of about

230 pm/mW is achieved. Note that the thermal tuning efficiency could be further improved if thermally isolating deep trenches are used.

**High-speed drivers and wavelength locking and tracking system**

Figure 3a shows the block diagram of the subsystems of the implemented transmitter chip. Each subsystem includes eight MRMs placed on an optical waveguide bus, electronic circuits for data generation and distribution, MRM drive and MRM wavelength control. The p-n section of each MRM is connected to a low-power high-speed modulator driver. To concurrently test all 32 channels, transmitting data is generated on-chip using pseudo-random bit sequence (PRBS) generators. Each PRBS generator can simultaneously provide data to 8 modulator drivers.

The dual sensing, actuation, and memory (DSAM) unit is used to tune and control the resonance wavelength of the MRMs. The DSAM unit of an MRM has two actuation channels. One is used to capacitively tune the MRM and the other one, in case needed, is used for thermal tuning of the MRM. A shared central control unit (which includes a digital ring selection circuit, a comparator and a current-to-voltage converter) is used to sequentially select and control the DSAM units. On-chip PRBS generators, synchronized with an off-chip clock, followed by serializer units, are used to generate the input NRZ bit-streams. The design of the on-chip bit-stream generator, which utilizes the standard PRBS-7 pattern[33], is similar to our 256Gb/s transmitter design[12]. Each PRBS generator creates a pseudo random bit-sequence at 32 Gb/s, which is seamlessly repeated, ensuring concurrent independent modulation of all 32 channels. The output of the PRBS generator is routed to the high-speed electrical data-path. Figure 3b shows the schematic of the electrical data-path, where the input bitstream is amplified using a series of inverter stages and then level-shifted. Within the level shifter, the data is routed to two signal paths, where a network of resistors and capacitors in each path is used to high-pass filter the signal and adjust its DC level.

The signals within each path are further amplified using inverters, with an additional carefully sized inverter stage on one of the paths to convert the signal from single-ended to differential while minimizing the delay mismatch. The level shifter outputs are routed to the MRM driver stage, which is formed using vertically stacked transistors, effectively increasing the output switch across the p-n junction of the MRM to about two times the supply voltage without exceeding the breakdown voltage of the transistors[34]. The driver output drives the p-n junction of the MRM differentially at a swing of about four times the DC supply voltage with no need for a negative DC supply. The series capacitor followed by the shunt resistor placed between the driver output and the MRM allows for adjusting the DC bias across the p-n junction independent of the driver output signal. The electrical path has a low cut-off frequency of about 7 MHz enabling the chip to operate at data-rates higher than 1 Gb/s when PRBS-7 scheme is used.

The measured and simulated normalized electro-optic response of the high-speed data path driving the MRM is shown in Fig. 3c. The entire high-speed data path (including the pre-amplifiers placed after the PRBS generators) has a bandwidth of about 17 GHz, which is in close agreement with simulation results.

The block diagram of the DSAM unit of an MRM is shown in Fig. 3d, where in the first actuation channel, the digital output of a 5-bit up/down counter is level-shifted, amplified using inverters (increasing the digital high logic from VDD to 2VDD), converted to an analog voltage, Vcap, using a 5-bit R-2R resistive ladder digital-to-analog converter (DAC) and then used to tune the capacitive section of the MRM. In the second actuation channel, the digital output of a 7-bit shift register is converted to an analog voltage using a 7-bit R-2R resistive ladder DAC and is used to tune the MRM heater. To monitor the MRM tuning and locking a SiGe photodiode (with a responsivity of about 0.9 A/W) is used to photo-detect 2% of the MRM output, which is sampled using a directional coupler. The design of the DSAM units follows our prior work[12]. DSAM units are activated sequentially (one at a time) using an on-chip digitally controlled selector, which is implemented using a shift register. The block diagram of the shared central control unit used for wavelength locking and carrier tracking is shown in Fig. 3e. The central control unit sequentially selects the MRMs. When an MRM is selected (by activating its enable signal, En), the feedback photo-current from the corresponding DSAM unit is routed to the central control unit, converted to a voltage, Vfb, (using an on-chip resistor or an off-chip potentiometer) and compared with a user-defined threshold. The threshold voltage, Vth, is set using a serially programed 6-bit shift register followed by an R-2R resistive ladder DAC. The threshold voltage sets the desired offset between the wavelength of each carrier and the resonance wavelength of the corresponding MRM. By comparing Vfb and Vth, the output of the comparator sets the up/down signal of the 5-bit counter within the DSAM unit, which counts up or down on the rising edge of the off-chip clock (Counter clk).

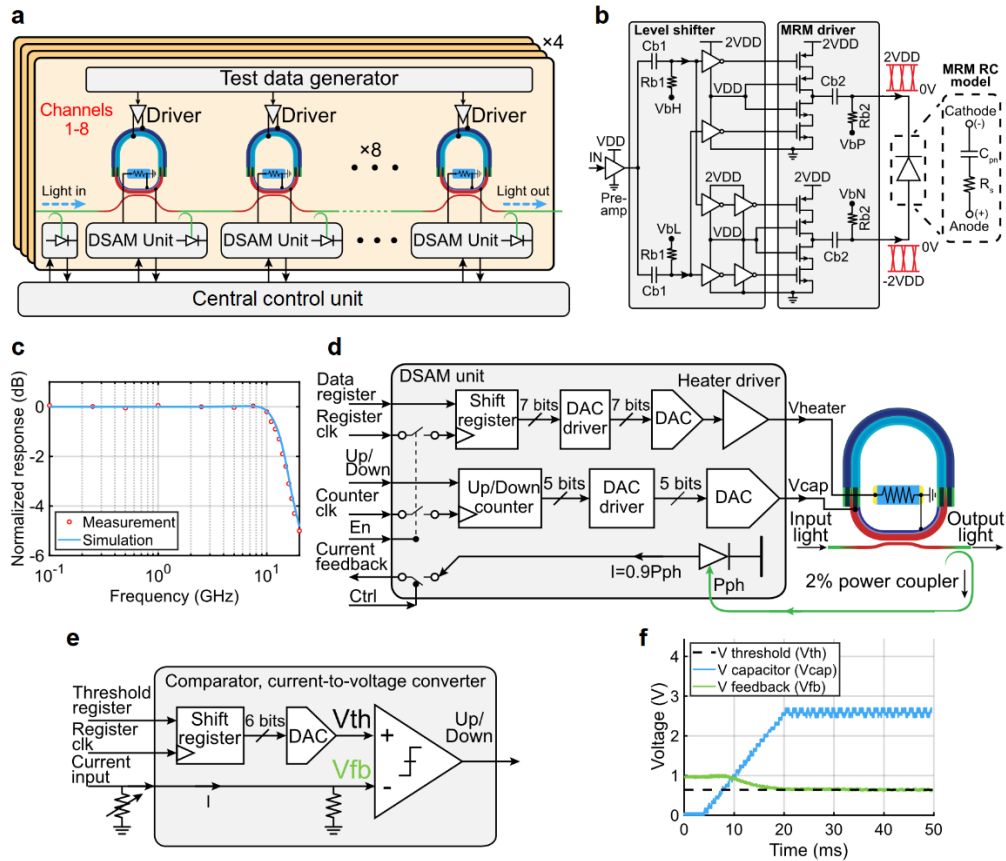

**Fig. 3 | MRM drive, tuning and locking**. **a,** Block diagram of the integrated 8-channel subsystem with eight MRMs on a bus, the data-generation, driver electronics and MRM tuning and locking control circuits. **b,** Schematic of the electrical data path including the pre-amplifier, level shifters and driver. **c,** Measured electro-optical response of the MRM. **d,** Diagram of the dual sensing actuation and memory (DSAM) unit, which controls the capacitive and heater sections of the MRM. **e,** Block diagram of the shared MRM locking and control circuit. **f,** Measured MRM capacitive locking process.

The measured MRM capacitive locking process is shown in Fig. 3f, where in a feedback loop, the voltage across the capacitive section is dynamically changed to match the feedback voltage to the threshold voltage, locking the resonance wavelength of the selected MRM to the wavelength of the corresponding target carrier. Note that DSAM units have a µW-level power consumption in the idle mode and when selected, they consume about 50 µW during the dynamic locking process.

**System measurement results**

Figure 4a shows the micrograph of the implemented 32-channel optical transmitter chip fabricated using GlobalFoundries GF45CLO CMOS SOI process.

To perform concurrent measurements of all 32 channels of the transmitter chip, a 16-channel laser source, formed using two Thorlabs PRO8000 units each housing 8 distributed feedback (DFB) laser modules, was used. The measured optical spectrum of the combined output of the 16-channel laser source, operating on the 1.6 nm international telecommunication union (ITU) grid, is shown in Fig. 4b, where a peak power variation of less than 0.5 dB across all carriers is observed.

Figure 4c shows the measurement setup, where the output of the 16-channel laser source is amplified, polarization adjusted and coupled into the chip through the input inverse designed grating coupler. The PRBS7 bit-stream is generated and distributed on-chip to simultaneously drive all 2-section MRMs. The clock signal for the on-chip

PRBS generator is provided using an off-chip frequency synthesizer. A microcontroller programs the on-chip wavelength locking and carrier tracking circuit.

Modulated carriers are coupled out of the chip and into the single-mode fibers using inverse designed grating couplers. To monitor each modulated carrier, a narrowband optical filter (Santec OTFC-910) with an insertion loss of approximately 4 dB is utilized before detection. To compensate for the coupling and filter losses, the output signal of the optical filter is amplified using an erbium-doped fiber amplifier (Thorlabs EDFA100S) and coupled to the GTRAN-GT40 optical receiver with differential outputs. One output of the receiver is connected to an Infiniium DCA-J 86100C sampling oscilloscope to monitor the output eye diagram, and the other output is used to measure the bit-error-rate (BER) using an SHF 11100A error analyzer.

Figure 4d shows the measured optical transmission response of the 32-channel transmitter chip for 4 buses (with 8 channels per bus) after sequential wavelength locking and tracking is engaged, aligning all channels to the corresponding input optical carriers. Over many measurements, majority of the tuning and tracking were performed using capacitive tuning with zero static power consumption. However, for certain cases, partial thermal alignment for a fraction of channels was needed (for which, an average power consumption of about 3.6 mW for channels with thermal tuning was measured).

Figure 5a shows the measured photo-detected output eye diagrams of all 32 channels operating simultaneously at 32 Gb/s/channel, corresponding to an aggregate data-rate of 1.024 Tb/s. In this measurement, while all channels are active, the narrow tunable optical band-pass filter placed after the chip output is tuned to monitor eye diagrams of all 32 channels (one at a time). Note that for the eye-diagram measurements, the total optical power before the chip, corresponding to the sum of the power of all 16 carriers, is about 21 dBm, which is limited by the power handling of the chip passivation layer. Higher optical power levels irreversibly damage the passivation over the input grating coupler. Figure 5b and 5c show the measured BER at 32Gb/s for different optical power levels at the modulator input and different per-channel data-rates, respectively, indicating that error-free operation (BER<$10^{-12}$) at 32 Gb/s requires optical power levels of 1 dBm or higher at the modulator input. Note that here, the maximum per-channel data-rate is mostly limited by the speed of the integrated electronics. Figure 5d shows the energy efficiency of the modulator for different per-channel data-rates, achieving 106 fJ/bit at 32 Gb/s and 32 fJ/bit at 5 Gb/s (based on $CV^2/4$ calculations[35]).

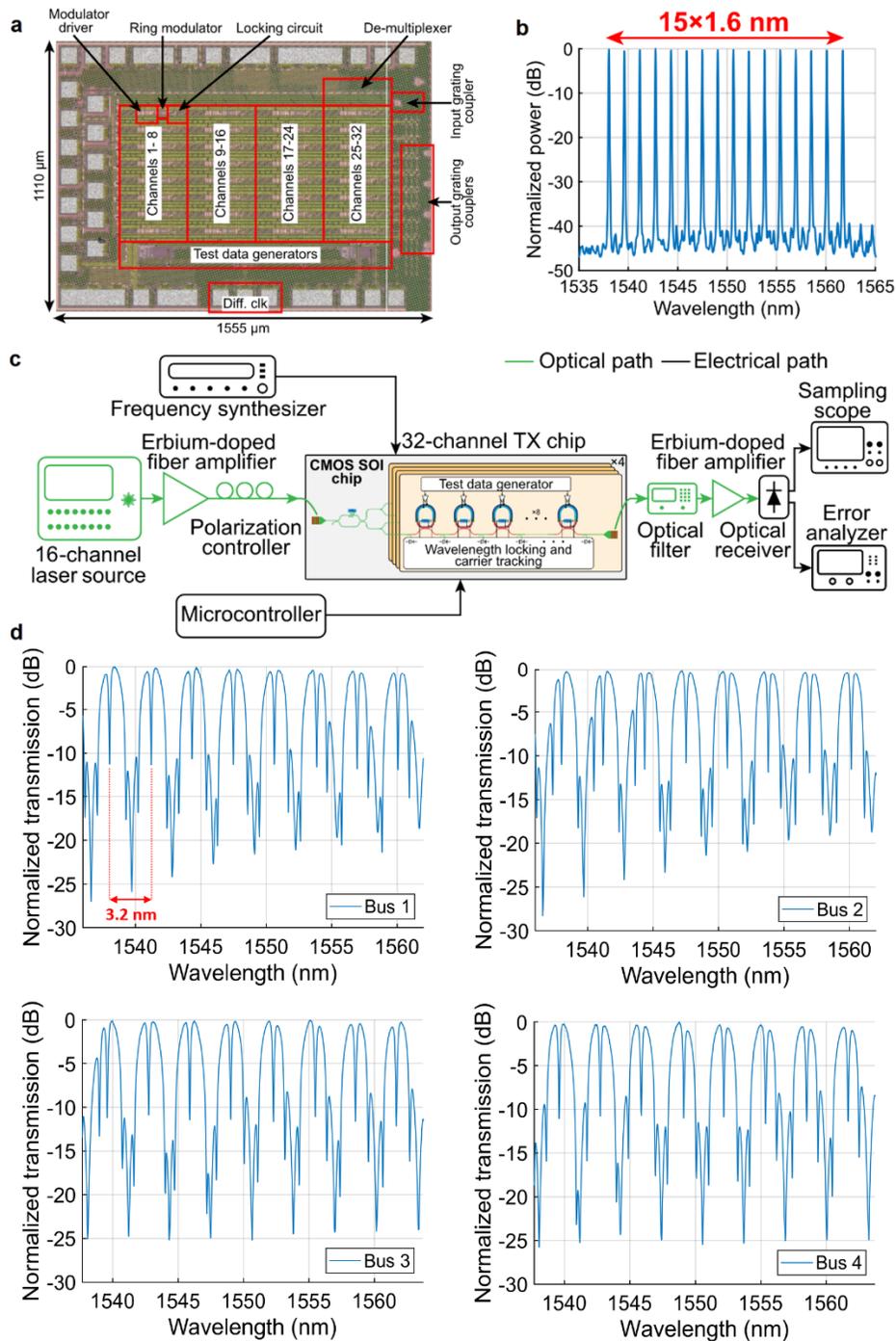

**Fig. 4 | Measurement setup and optical response of different channels. a,** Micrograph of the implemented 32-channel 1.024 Tb/s transmitter chip. **b,** Measured optical spectrum of the input 16-channel laser source, serving as 16 optical carriers with a carrier-to-carrier spacing of 1.6 nm. **c,** Block diagram of the measurement setup for the 32-channel optical transmitter chip. **d,** Measured normalized optical transmission responses of the 32-channel transmitter chip for buses 1-4 with a 3.2 nm spacing between consecutive resonance wavelengths.

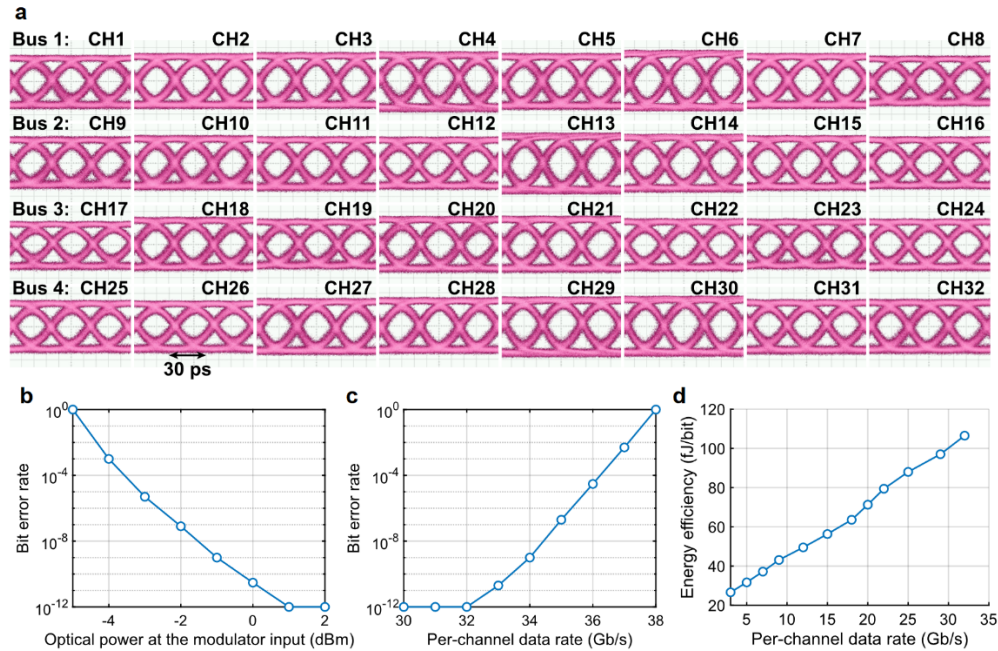

**Fig. 5 | Chip measurement results**. **a,** Measured eye diagrams of all 32 channels working simultaneously at per-channel data-rate of 32 Gb/s, corresponding to an aggregate data-rate of 1.024 Tb/s. **b,** Measured BER versus input optical power at the modulator input for per-channel data-rate of 32 Gb/s. **c,** Measured BER versus per-channel data-rate. **d,** Energy efficiency of the modulator versus per-channel data-rate.

## Discussion

In summary, we have demonstrated an energy efficient single chip 32-channel optical transmitter based on wavelength-division multiplexing. Utilizing monolithic integration approach, we designed a system with inverse design input/output grating couplers, 2-section p-n-capacitive MRMs, MZI based demultiplexer, energy efficient modulator drivers, wavelength locking and carrier tracking circuits and test data generator units. The capacitively tuned MRM offers wavelength locking and carrier tracking at zero static power consumption. The maximum per-channel error-free data-rate of 32 Gb/s was achieved. The chip supports simultaneous modulation of 16 carriers within 32 channels, resulting in an aggregate data-rate of up to 1.024 Tb/s. The system is implemented using GlobalFoundries 45CLO, a 45 nm CMOS SOI process, without any process modification or post-processing. Monolithic integration results in reduced parasitics in the interface between the electrical drivers and MRMs. As a result, the modulation efficiency reaches 32 fJ/b at 5 Gb/s and 106 fJ/b at 32 Gb/s, which includes the power consumption of the on-chip autonomous MRM capacitive wavelength tuning and locking. Furthermore, a bandwidth density of 8.0 Tb/s/mm$^2$ for the modulators, drivers, and control units is achieved. The performance of the implemented 32-channel transmitter system is compared with that of previously published state-of-the-art multi-channel NRZ optical transmitter systems in Extended Data Table 1. This work simultaneously achieves the highest reported aggregate data-rate, bandwidth density and energy efficiency, and advances the state of the optical interconnects for future data centers and AI applications.

**Methods**

**Chip fabrication**

The transmitter chip was monolithically integrated using the GlobalFoundries 45CLO CMOS-SOI silicon photonic platform. This technology supports devices with an $f_T$ of up to 280 GHz, making it well-suited for RF signal generation and conditioning on the same chip alongside the optical devices and corresponding control electronics. The monolithic integration enables the realization of optical transceivers with reduced packaging complexity and less parasitics.

The design incorporates photonic waveguides with an optical loss of approximately 1.4 dB/cm at around 1550 nm and photodiodes exhibiting a responsivity of approximately 0.9 A/W and bandwidth of about 50 GHz.

**System wavelength planning and carrier demultiplexing**

The relative wavelength response of the MRMs and carriers is designed to achieve high energy efficiency while maintaining a low crosstalk. Extended Data Fig. 1 shows the effect of the FSR of the MRMs on crosstalk for a given carrier-to-carrier spacing. Extended Data Fig. 1a shows a practical example, where 16 carriers (labeled from 1 to 16) with a carrier-to-carrier spacing of $\lambda_{spacing}$, and an MRM (whose transmission response is shown with dashed red lines) with an FSR larger than $\lambda_{spacing}$ but smaller than the input optical bandwidth are used. In this case, when one of the resonances of the MRM aligns with the target optical carrier, another resonance wavelength of the MRM could fall close to a non-target carrier (say $\Delta\lambda_c$ away) causing crosstalk. Note that, the crosstalk is minimized when $\frac{\Delta\lambda_c}{\lambda_{spacing}}$ is maximized. Extended Data Fig. **1b** shows $\frac{\Delta\lambda_c}{\lambda_{spacing}}$ as a function of $\frac{FSR}{\lambda_{spacing}}$ for 16 carrier wavelengths, where the minimum achievable crosstalk (i.e. for the largest possible $\Delta\lambda_c$ of $0.5\lambda_{spacing}$ or equivalently $\frac{\Delta\lambda_c}{\lambda_{spacing}} = 0.5$) occurs for certain values of $\frac{FSR}{\lambda_{spacing}}$ (i.e. 8.5, 9.5, 10.5, …). In this case, given the desired $\lambda_{spacing}$, the circumference of the MRM could be set to achieve the required FSR minimizing the crosstalk. Note that since a larger MRM often exhibits a higher quality factor and is less sensitive to fabrication process variations, a reasonable design choice would be to select the largest MRM that achieves $\frac{\Delta\lambda_c}{\lambda_{spacing}} = 0.5$ (which in this example would be an MRM with an FSR of $8.5\lambda_{spacing}$).

Crosstalk can be further improved if instead of using consecutive carriers, odd/even carrier de-multiplexing technique[26] is used (Extended Data Fig. **2c**). In this case, only every other optical carrier (odd or even) with spacing of $2\lambda_{spacing}$ between adjacent carriers will be passed to MRMs on a bus. Extended Data Fig. **1d** shows $\frac{\Delta\lambda_c}{\lambda_{spacing}}$ as a function of $\frac{FSR}{\lambda_{spacing}}$ for the odd/even carrier multiplexing case, where the minimum achievable crosstalk occurs for certain values of $\frac{FSR}{\lambda_{spacing}}$ (i.e. 9, 11, 13, ...). Here, the largest possible $\Delta\lambda_c$ of $1\lambda_{spacing}$ results in the minimum crosstalk, indicating that, for the same number of optical carriers in the system, the use of odd/even wavelengths de-multiplexing technique further reduces the crosstalk.

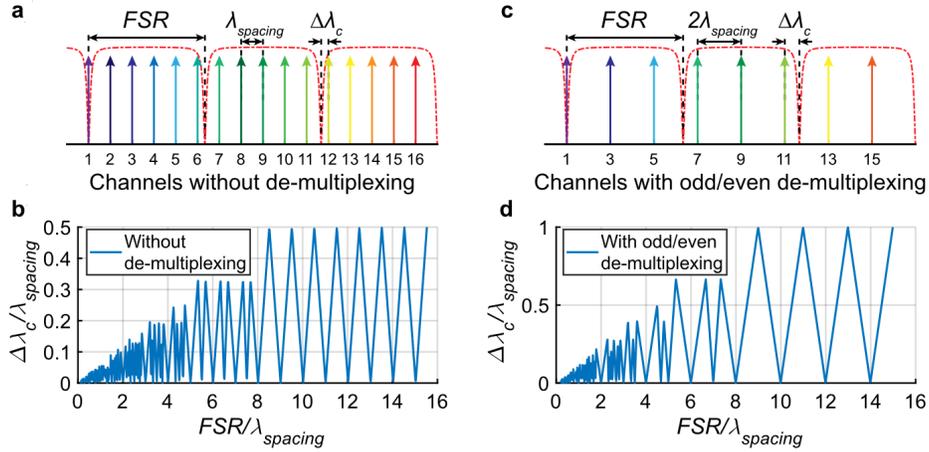

**Extended Data Fig. 1 | System wavelength planning and carrier demultiplexing.**
**a,** System wavelength configuration with 16 carriers and an MRM with an FSR larger than carrier-to-carrier spacing but smaller than the input optical bandwidth and **b,** the corresponding minimum achievable crosstalk condition for different FSR of the MRM normalized to carrier-to-carrier spacing. **c,** System wavelength configuration with odd/even carrier de-multiplexing and **d,** the corresponding minimum achievable crosstalk condition for different FSR of the MRM normalized to carrier-to-carrier spacing.

**Micro-ring modulator design parameters**

The intensity transmission function of a ring resonator is given by[31]

$$T = \frac{I_{pass}}{I_{input}} = \frac{2 - a^2 - k^2 - 2\sqrt{1-a^2}\sqrt{1-k^2}\cos(\beta L)}{1 + (1-a^2)(1-k^2) - 2\sqrt{1-a^2}\sqrt{1-k^2}\cos(\beta L)}, \quad (1)$$

where $a^2$, $k^2$, $\beta = 2\pi n/\lambda$, $n$, $\lambda$ and $L$ are the power loss in the ring per round-trip ( including both propagation loss in the ring and the loss of the couplers), the power coupling coefficient to the ring from the bus waveguide, the propagation constant of the circulating mode, the waveguide refractive index, the operation wavelength and the circumference of the ring, respectively. In this case, the extinction ratio (ER), quality factor (Q), and modulation efficiency (ME) are defined as

$$ER = 10 \log_{10} \left(\frac{T_{max}}{T_{min}}\right), \quad (2)$$

$$Q = \frac{\lambda_{res}}{FWHM}, \quad (3)$$

$$ME = \frac{dT}{d\lambda}, \quad (4)$$

where $T_{max}$ and $T_{min}$ represent the maximum and minimum transmission coefficients and $FWHM$ is the full-width at half-maximum value of the resonance intensity spectrum that can be derived using the intensity transmission function in Eq. 1.

Assuming $n = 4.0$, the ER and Q versus power coupling coefficient for different loss values inside a ring resonator with $L = 47$ μm are calculated and depicted in Extended Data Fig. 2a. While the ER peaks for the case that the ring resonator is critically coupled, Q drops monotonically as coupling coefficient increases. Extended Data Fig. 2b shows the ME versus power coupling coefficient. Note that the maximum of ME occurs when the ring resonator is slightly under-coupled. For example, for the case that the ring loss is 13% (dashed lines in Extended Data Figs. 2a and 2b), the critical coupling (where the ER is maximum) occurs at power coupling coefficient of 0.13, while the ME is maximum at power coupling coefficient of around 0.07 (corresponding to an ER of about 10 dB). These trade-offs were used to design efficient micro-ring-based modulators for the system.

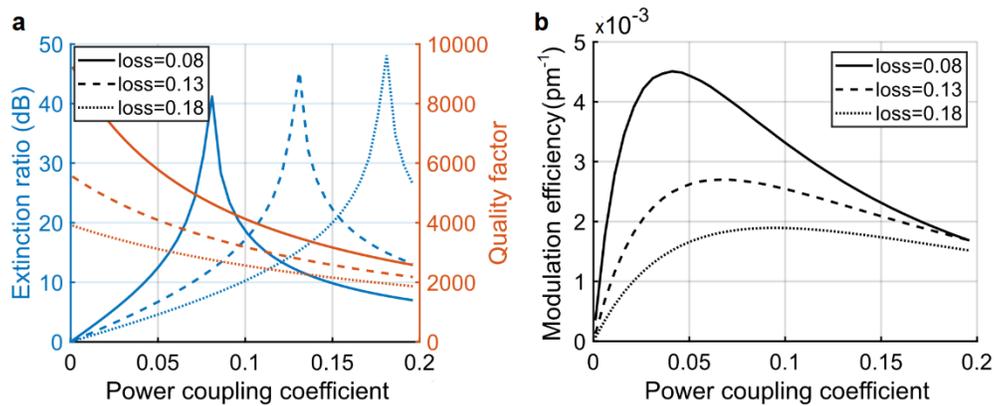

**Extended Data Fig. 2 | Micro-ring resonator design parameters**. **a,** Calculated extinction ratio and quality factor of the ring resonator versus power coupling coefficient for different loss values in the ring. **b,** Calculated modulation efficiency for the ring resonator versus power coupling coefficient for different loss values in the ring. Here, $n = 4.0$ and $L = 47$ μm.

### Acknowledgement
This work was supported by Defense Advanced Research Projects Agency PIPES program under contract number HR0011-19-2-0016.

### Author contributions
K.O., J.V. and F.A. conceived the project idea. K.O. led the overall chip and PCB design and conducted full system measurements in collaboration with A.S. A.P. assisted with the architecture design and designed the broadband Mach–Zehnder interferometers as well as various control electronics for wavelength locking. H.H. developed the on-chip pattern generator. G.H.A. designed and characterized the inverse-designed input/output grating couplers. J.V. directed and supervised the design, implementation and characterization of inverse designed devices. All authors contributed to writing the manuscript. F.A. directed and supervised the project.

### Disclosures
The authors declare no conflicts of interest.

### Data availability
Data underlying the results presented in this paper are not publicly available at this time but may be obtained from the authors upon reasonable request.

|  | **Ref. 36** | **Ref. 37** | **Ref. 38** | **Ref. 39** | **This work** |
|---|---|---|---|---|---|
| **Integration approach** | Monolithic | Hybrid (flip-chip) | Hybrid (flip-chip) | Hybrid (flip-chip) | Monolithic |
| **Modulator type** | Single section p-n based micro-ring | Single section p-n based micro-disk | Single section p-n based micro-disk | Single section p-n based micro-ring | 2-section p-n and capacitive based micro-ring |
| **Operation band** | C | O | C | O | C |
| **PIC technology** | Monolithic | Custom | Custom | Custom | Monolithic |
| **EIC technology** | 45 nm CMOS SOI | 12 nm FinFET | 28 nm CMOS | 28 nm CMOS | 45 nm CMOS SOI |
| **Post-processing** | Yes | No | No | No | No |
| **ER** | 10 dB | 7 dB | 9 dB | 5 dB | 11 dB[(1)] |
| **Wavelength tuning/locking** | Thermal tuning and locking | Thermal tuning and locking | No | Thermal tuning and locking | **Capacitive and thermal tuning/locking** |
| **Tuning efficiency** | 1.25 nm/mW | 0.27 nm/mW | Not available | Not reported | **Capacitive: 0.045 nm/V at zero power. Thermal: 0.23 nm/mW** |
| **Aggregate data-rate** | 55 Gb/s | 372 Gb/s | 800 Gb/s | 400 Gb/s | **1024 Gb/s** |
| **Concurrent aggregate data-rate** | 5 Gb/s | 12 Gb/s | 10 Gb/s | 400 Gb/s | **1024 Gb/s** |
| **BER for Concurrent demonstration** | Not available | Not available | Not available ($10^{-8}$ to $10^{-10}$ after link Rx) | Not available | **$10^{-12}$** |
| **Number of laser/carriers** | 1 (one channel at a time) | 1 (one channel at a time) | 4 | 8 | **16** |
| **Number of channels** | 11 | 31 | 80 | 8 | 32 |
| **Bandwidth density** | Not available | Not available | 5.3 Tb/s/mm$^2$ (excluding drivers) | Not available | **8.0 Tb/s/mm$^2$ (including MRM drivers)** |
| **Data-rate per channel** | 5 Gb/s | 12 Gb/s | 10 Gb/s | 50 Gb/s | Up to 32 Gb/s |
| **Modulation energy efficiency** | 30 fJ/b @ 5 Gb/s | 150 fJ/b @ 12 Gb/s [(2)] | 50 fJ/b @ 10 Gb/s plus 176 fJ/b for thermal control system | 0.56 pJ/b @ 50 Gb/s [(3)] | **32 fJ/b @ 5 Gb/s 45 fJ/b @ 10 Gb/s 106 fJ/b @ 32 Gb/s** |

[(1)] For $V_{cap}=0$.
[(2)] Includes data serialization.
[(3)] Includes data path.

**Extended Data Table. 1 | Performance summary and comparison with selected NRZ optical transmitters.**